\documentclass{webofc}
\usepackage[varg]{txfonts}   
\begin{document}
\vspace*{-0.5cm}
\hfill YITP-22-159
\vspace*{0.25cm}

\title{Study of Hadron Masses with Faddeev-Popov
Eigenmode Projection in the Coulomb Gauge}
%
%

\author{\firstname{Hiroki} \lastname{Ohata}\inst{1}\fnsep\thanks{\email{hiroki.ohata@yukawa.kyoto-u.ac.jp}} \and
        \firstname{Hideo} \lastname{Suganuma}\inst{2}\fnsep\thanks{\email{suganuma@scphys.kyoto-u.ac.jp}} 
}

\institute{Yukawa Institute for Theoretical Physics, Kyoto University, Kyoto 606-8502, Japan
\and
Department of Physics, Kyoto University, 
Kitashirakawa-oiwake, Sakyo, Kyoto 606-8502, Japan
          }

\abstract{%
Using SU(3) lattice QCD, we investigate role of spatial gluons for hadron masses in the Coulomb gauge, considering the relation between QCD and the quark model. 
From the Coulomb-gauge configurations at the quenched level on a $16^3 \times 32$ lattice at $\beta$ = 6.0, we  
consider the $\vec{A} = 0$ projection, 
where all the spatial gluon fields are set to zero.
In this projection, the inter-quark potential is unchanged. 
We investigate light hadron masses and find that nucleon and delta baryon masses are almost degenerate. This result suggests that 
the N-$\Delta$ mass difference arises from the color-magnetic interactions, which is consistent with the quark model picture.
Next, as a generalization of this projection, 
we expand spatial gluon fields in terms of Faddeev-Popov eigenmodes 
and leave only some partial components.
We find that the ${\rm N}-\Delta$ and $0^{++}-2^{++}$ glueball mass splittings are almost reproduced only with 1 \% low-lying components. 
This suggests that low-lying color-magnetic interaction leads 
to the hadron mass splitting. 
}
\maketitle

\section{Introduction}

In spite of the great success of the quark model~\cite{DeRujula:1975qlm}, 
its connection with quantum chromodynamics (QCD) is not yet clear, 
and to clarify the relation between these theories is one of the most important problems in hadron physics.
A significant difference between them is the main dynamical degrees of freedom: while QCD is formulated with current quarks and gluons, the quark model is described only with the massive  constituent quarks.
The other significant difference would be the symmetries they possess.
QCD has local color $\mathrm{SU}(3)$ symmetry, whereas the quark model only has global color $\mathrm{SU}(3)$ symmetry reflecting the absence of dynamical gluons.
This suggests that the quark model is a low-energy effective theory of QCD in some gauge.

Then, what gauge corresponds to the quark model?
Here, we need not respect full Lorentz symmetry, 
since the nonrelativistic quark model has only the spatial-rotation symmetry. 
We consider that gauge to be the Coulomb gauge.
There are several reasons for this choice.
First, the Coulomb gauge leaves global color $\mathrm{SU}(3)$ symmetry at each time slice, which might correspond to the global color symmetry in the quark model.
Second, the Coulomb gauge is globally defined by minimizing spatial gauge-field fluctuations,  
and therefore one expects only a small fluctuation 
of the spatial gluon field, which might explain the absence of dynamical gluons in the quark model.
Third, in the Coulomb gauge, the temporal gauge field $A_0$ 
is no more dynamical but appears as a potential for quarks, 
which might give the static potential in the quark model. 
Several researchers have already studied the relation between the quark model and Coulomb gauge QCD~\cite{Szczepaniak:2001rg}.

We investigate the relation between the quark model and Coulomb gauge QCD using gauge configurations 
generated in SU(3) lattice QCD.
The organization of this paper is as follows.
In Sec.~\ref{sec:Coulombgauge}, we briefly review 
QCD in the Coulomb gauge.
In Sec.~\ref{sec:A0projection}, we propose 
$\vec{A} = 0$ projection, i.e., 
removal of spatial gluons from lattice gauge configurations in the Coulomb gauge, 
and investigate the static inter-quark potential and hadron masses. 
In Sec.~\ref{sec:eigenmodeprojection}, we propose a generalization of the $\vec{A} = 0$ projection in the Coulomb gauge utilizing the Faddeev-Popov eigenmodes.
In Sec.~\ref{sec:hadron_mass}, we apply this  projection to light hadron and glueball masses. 
Section~\ref{sec:summary} is devoted to Summary and Conclusion.
This paper is based on Ref.~\cite{Ohata:2022iqm}.

\section{Coulomb gauge QCD} \label{sec:Coulombgauge}
We briefly review Coulomb gauge QCD, which 
is locally defined as
$\partial_i A_{s,i}^a = 0$ for the gluon field $A_{s,\mu}=A_{s,\mu}^aT^a \in {\rm su}(N_c)$. 
In the Coulomb gauge, 
the spatial gluon fields $A_{i}^a$ are canonical variables and behave as dynamical variables, 
whereas the temporal gluon field $A_0^a$ 
just behaves as a potential field.
The global definition of the Coulomb gauge is to minimize 
\begin{equation}
R_{\mathrm{C}}[A] \coloneqq \int dt~d^3s \sum_{i,a}~ [A_{s,i}^a(t)]^2
\end{equation}
under the gauge transformation, 
which satisfies the local Coulomb gauge condition.
In fact, the gauge fluctuation of the spatial gluon field is strongly suppressed in the Coulomb gauge. 

Thus, in the Coulomb gauge, 
spatial gluons are forced to be minimized and 
the temporal gluon just provides a potential. 
This situation might be preferable to match 
QCD and the quark model,
since the quark model only has quark degrees of freedom and an inter-quark potential.

In the path-integral formalism, the generating functional of the Yang-Mills (YM) theory 
in the Coulomb gauge is expressed by
\begin{equation}
Z = \int DA\, e^{iS[A]}~\delta(\partial_i A_i^a)~{\rm Det}(D_i\partial_i)
\end{equation}
with the YM action $S[A]$.
There appears the Faddeev-Popov (FP) operator, 
\begin{equation}
M^{ab} \coloneqq D_i^{ab}\partial_i =\partial_i^2\delta^{ab}+gf^{abc}A^c_i \partial_i, 
\label{eq:continuumFP}
\end{equation}
as a weight of the gauge orbital in the Coulomb gauge.
This FP operator is one of the key operators in the Coulomb gauge, 
and its inverse gives the propagator of the FP ghost field.

In lattice QCD, the gauge degrees of freedom is described by the link variable 
$U_{s,\mu} \coloneqq e^{iagA_{s,\mu}} \in {\rm SU}(N_c)$, where $g$ is the gauge coupling and $a$ the lattice spacing, and the global definition of the Coulomb gauge is to maximize 
\begin{equation}
R_{\mathrm{C}} [U] \coloneqq \sum_{t,s,i} {\rm Re}~ 
{\rm Tr} U_{s,i}(t) \label{eq:Coulomb_def}
\end{equation}
under the gauge transformation.
This global definition leads to the local gauge condition
$\partial^B_i {\cal A}_{s,i} = 0$,  
where $\partial^B$ denotes the backward derivative and 
\begin{equation}
{\cal A}_{s,i} \coloneqq  \frac{1}{2iag}\{U_{s,i}-U_{s,i}^\dagger\}\bigg{|}_{\rm traceless} \in {\rm su}(N_c)
\end{equation}
goes to the original gluon field $A_i(s)$ in the continuum limit.

In lattice QCD in the Coulomb gauge, the generating functional of the gauge sector is written by
\begin{equation}
Z= \int DU\, e^{-S[U]}~ \delta(\partial_i {\cal A}_i^a)~ {\rm Det}(M),
\end{equation}
where the FP operator in the Coulomb gauge takes the form~\cite{Iritani:2012bc} of 
\begin{equation}
M_{x,y}^{a,b} = \sum_{i} A_{x,i}^{a,b} \delta_{x,y} - 
B_{x,i}^{a,b} \delta_{x + \hat{i}, y} - C_{x,i}^{a,b} \delta_{x - \hat{i}, y},
\label{eq:FPlattice}
\end{equation}
with
\begin{align}
\!\!\!\! A_{x,i}^{a,b} = {\rm Re} {\rm Tr}[\{T^a, T^b\} ( U_{x,i} + U_{x - \hat{i},i})],~
B_{x,i}^{a,b} = 2 {\rm Re} {\rm Tr}[T^b T^a U_{x,i}],~
C_{x,i}^{a,b} = 2 {\rm Re} {\rm Tr}[T^a T^b U_{x - \hat{i},i}],\!
\end{align}
at each time slice, 
which leads to Eq.~(\ref{eq:continuumFP}) in the continuum limit.
Note that $M$ is a real symmetric matrix, i.e.,  
$M^{a,b}_{x,y}=M^{b,a}_{y,x} \in {\bf R}$, 
and hence its eigenvalues $\lambda_n$ are real.

In the Coulomb gauge, this FP operator is a key quantity to control the gauge-orbital weight, 
and has been studied in the context of 
the Gribov horizon~\cite{Gribov:1977wm,Zwanziger:1995cv}, 
the instantaneous color-Coulomb potential~\cite{Zwanziger:1995cv,Zwanziger:1998ez,Szczepaniak:2001rg,Zwanziger:2002sh},
and a confinement scenario of the gluon-chain picture~\cite{Greensite:2001nx,tHooft:2002pmx}.

\section{Lattice QCD simulation and $\vec{A} = 0$ projection} \label{sec:A0projection}
We perform SU(3) quenched lattice QCD calculation at $\beta = 6.0$ on a $L_s^3 \times L_t = 16^3 \times 32$ lattice.
We impose periodic boundary conditions for link variables in all directions.
We generate 500 gauge configurations, which are picked up every
$500$ sweeps after a thermalization of $5000$ sweeps.
For these gauge configurations, we perform Coulomb gauge fixing by maximizing Eq.~(\ref{eq:Coulomb_def}) under the SU(3) gauge transformation.
Thus, we obtain representatives of the QCD vacuum in the Coulomb gauge.

\begin{figure}[htb]
\begin{center}
\includegraphics[width=0.5\linewidth]{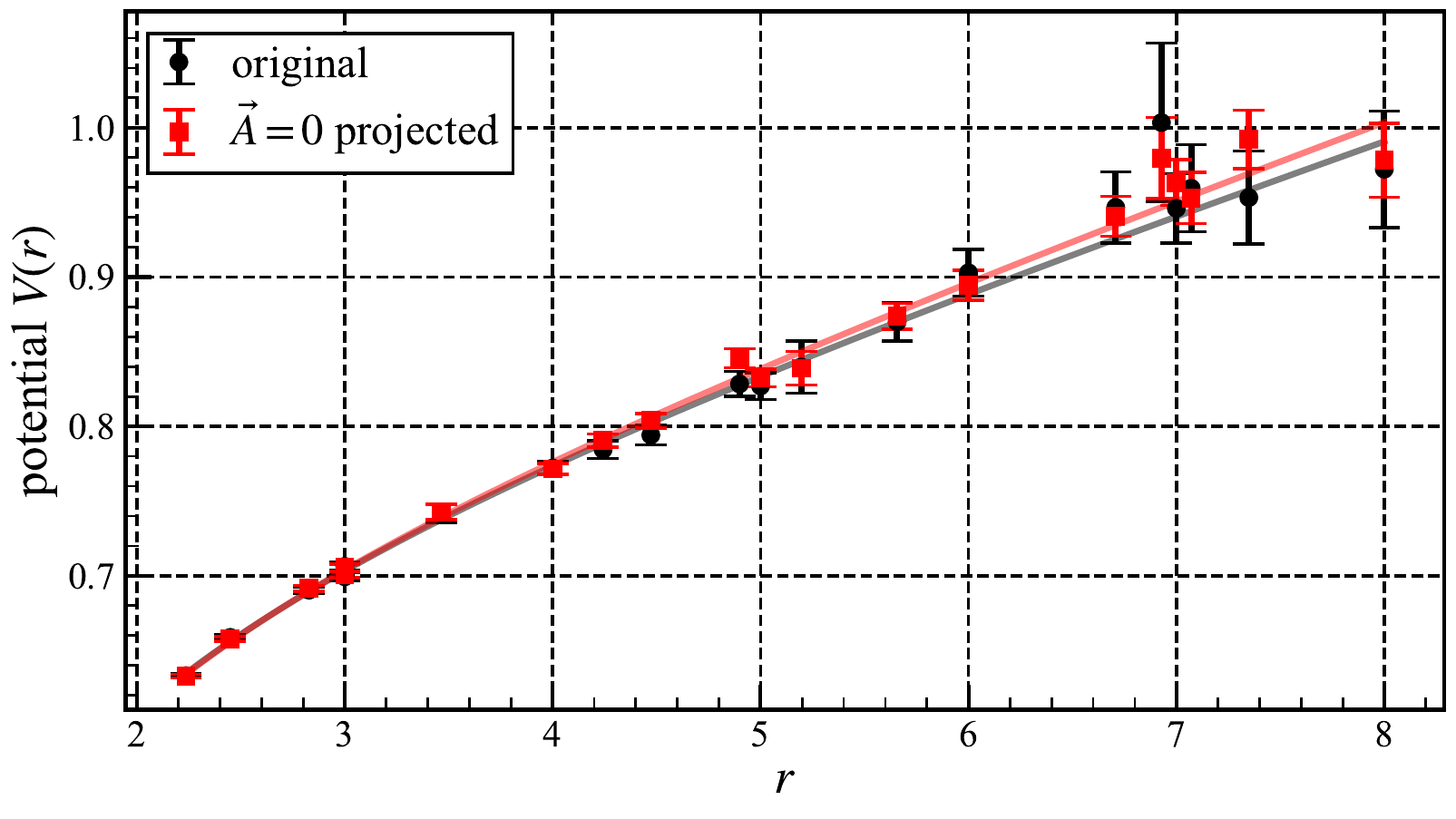}
\end{center}
\vspace{-0.6cm}
\caption{\label{fig:qqbar_pot}
The static quark-antiquark potentials evaluated from the original configurations (black) and $\vec{A} = 0$ projected ones in the Coulomb gauge (red).
The black and red curves are the fit results with the Cornell potential $V(r) = -A / r + \sigma r + \mathrm{const}$.
}
\end{figure}

The static inter-quark potential is one of the most fundamental constituents in the quark model.
In QCD, it can be calculated from correlator of the Polyakov loops.
Since the Polyakov loop is defined only by temporal gauge fields, the static quark-antiquark potential is independent of spatial gauge fields in the gauge configurations.
To see this, we define $\vec{A} = 0$ projection as a simple removal of spatial gluons in the generated gauge configurations in the Coulomb gauge, 
and apply it to the static quark-antiquark potential.
Figure~\ref{fig:qqbar_pot} shows the static  quark-antiquark potentials calculated from Wilson loops, evaluated from the original gauge configurations and $\vec{A} = 0$ projected ones in the Coulomb gauge. 
They agree very well in the whole distance region as expected.
In the quark model, the inter-quark potential is introduced by hand. In lattice QCD in the Coulomb gauge, we can keep the original static inter-quark potential by just leaving the temporal gauge fields unchanged.

Using the $\vec{A} = 0$ projection 
in the Coulomb gauge, 
we next investigate hadron masses 
to see the role of spatial gluons, 
which gives some hadron-mass splitting 
via the color-magnetic interactions
in the quark model.
Since hadron masses are not purely static quantities, their values under the $\vec{A} = 0$ projection are non-trivial.
We use the clover fermion ($\mathcal{O}(a)$ improved Wilson fermion) with a non-perturbatively determined clover coefficient $c_{\mathrm{SW}} = 1.769$~\cite{Luscher:1996ug}.
We consider the light hadrons of the pion, rho meson, nucleon, and delta baryon.
We extract pion mass $m_\pi$ and rho meson mass $m_\rho$ from correlators defined as
\begin{align}
C(t) = \sum_{\vec{x}} M^{\mathrm{ud}}(\vec{x}, t) M^{\mathrm{ud}}(\vec{0}, 0)^\dag, \quad 
M^{\mathrm{ud}} = 
\begin{cases}
\overline{u} \gamma_5 d & \text{for pion ($\pi^-$)}, \\
\overline{u} \gamma_i d & \text{for rho meson ($\rho^-$)}.
\end{cases}
\end{align}
For nucleon mass $m_{\mathrm{N}}$ and delta baryon mass $m_{\Delta}$, we use correlators
\begin{align}
\!\!
C(t) = \Gamma_{\alpha \beta} \sum_{\vec{x}} B^{\mathrm{udu}}_{\alpha}(\vec{x},t) \overline{B}_{\beta}^{\mathrm{udu}}(\vec{0},0), \quad
B^{\mathrm{udu}} = 
\begin{cases}
\epsilon^{abc} ((u^{\mathrm{T}})^a C \gamma_5 d^b) u^c \\ 
\qquad \qquad \qquad \text{for nucleon (proton)}, \\
\epsilon^{abc} \{
2((u^{\mathrm{T}})^a C \gamma_i d^b) u^c
+ ((u^{\mathrm{T}})^a C \gamma_i u^b) d^c
\} \!\!\!\! \\
\qquad \qquad \qquad \text{for delta baryon ($\Delta^+$)}.
\end{cases}
\end{align}
Here $C = \gamma_0 \gamma_2$ is the charge conjugation matrix, and we set $\Gamma = (1 \pm \gamma_0)$ to extract components proportional to these parity projection operators.

\begin{figure*}[htb]
\includegraphics[width=0.245\linewidth]{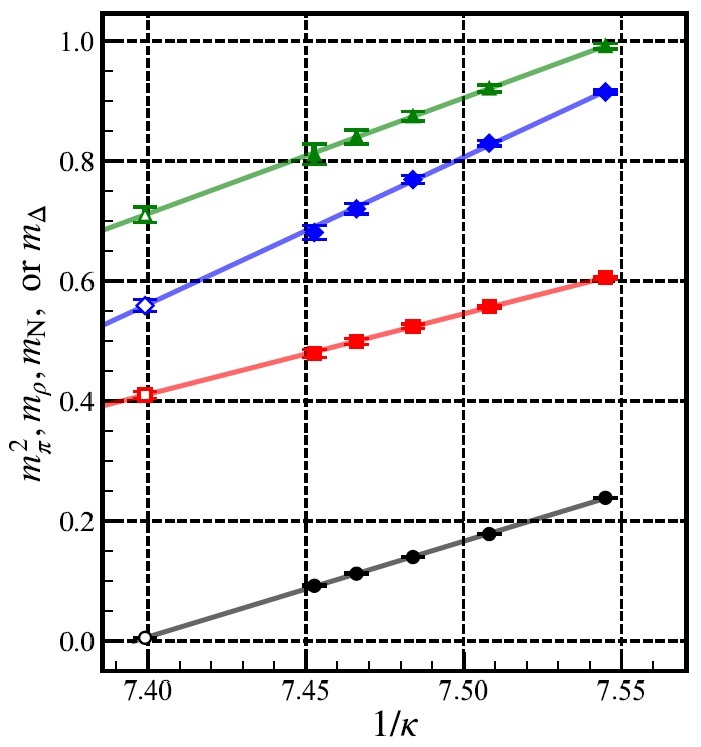}
\includegraphics[width=0.245\linewidth]{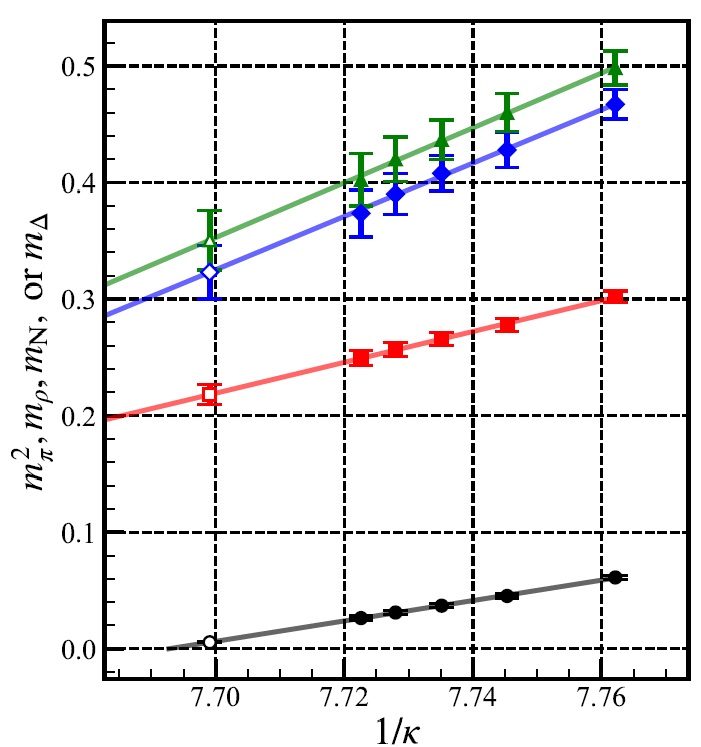}
\includegraphics[width=0.245\linewidth]{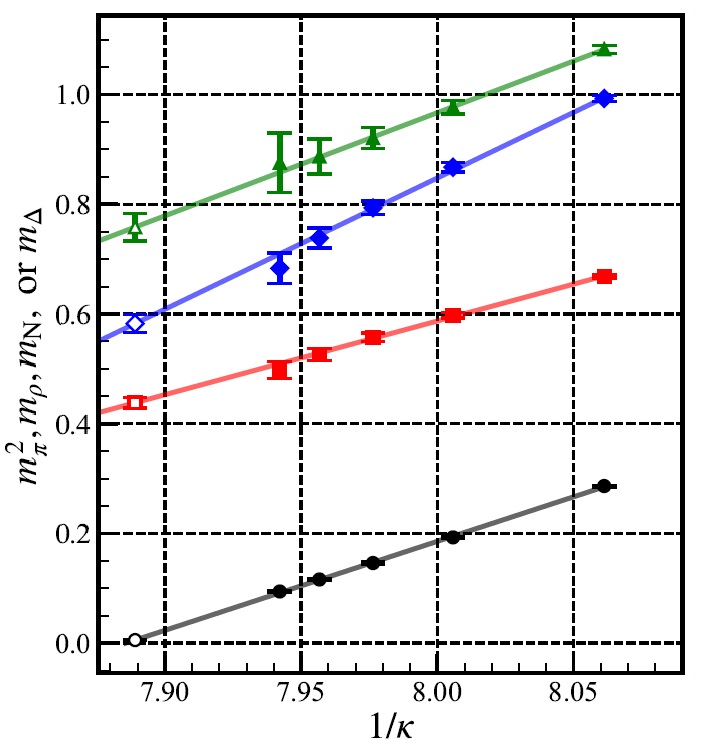}
\includegraphics[width=0.245\linewidth]{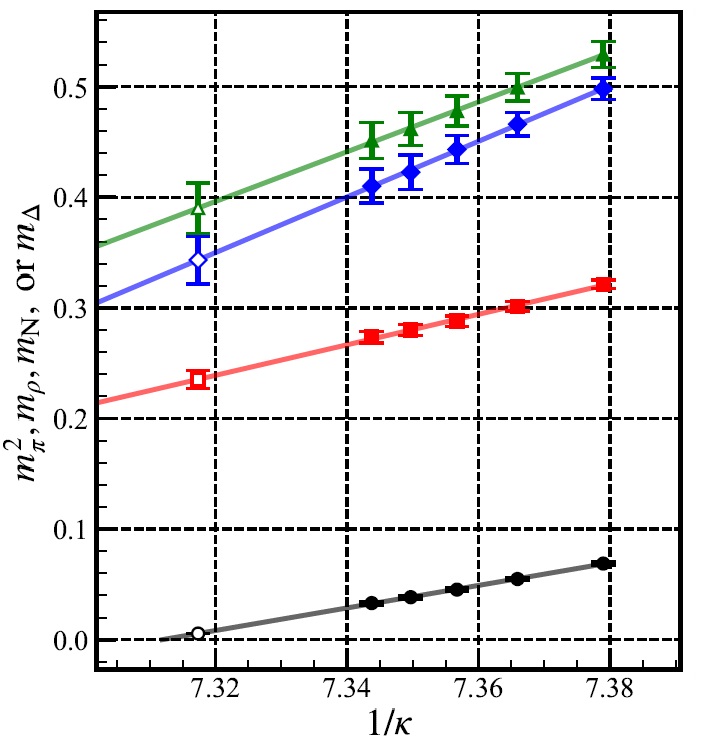}
\caption{\label{fig:hadron_masses}
Hadron mass measurement (a) from the original gauge configuration;
(b) under the $\vec A$ = 0 projection in the Coulomb gauge; 
(c) under the eigenmode projection with 5 \% low-lying FP eigenmodes;
(d) under the eigenmode projection with 95 \% high-lying FP eigenmodes.
}
\end{figure*}

Figure~\ref{fig:hadron_masses}~(a) shows light hadron mass measurement from the original gauge configuration.
For each hoppoing parameter $\kappa$ and sector, hadron mass is obtained from single cosh / exponential fit to the correlation function.
From linear chiral extrapolations of $m_{\pi}^2$ and $m_{\rho}$, 
the physical hopping parameter and the lattice spacing are determined so that $m_{\pi} = 0.14 {\rm GeV}$ and $m_{\rho} = 0.77 {\rm GeV}$.
The lattice spacing is found to be $a = 0.1051(15) {\rm fm}$.
Nucleon and delta baryon masses are found to be $m_{\mathrm{N}} = 1.050(24) {\rm GeV}$ and $m_{\Delta} = 1.334(31) {\rm GeV}$, 
which are consistent with a quenched lattice QCD simulation with a similar lattice setup~\cite{JLQCD:2002zto}.

Figure~\ref{fig:hadron_masses}~(b) shows light hadron mass measurement under the $\vec{A} = 0$ projection. 
Here we assume that the lattice spacing is the same as the one obtained from the original gauge configuration because the static quark-antiquark potential is unchanged.
The physical hopping parameter is determined so that $m_{\pi} = 0.14 {\rm GeV}$.
Rho meson, nucleon, and delta baryon masses are found to be $m_{\rho} = 0.409(17) {\rm GeV}, m_{\mathrm{N}} = 0.607(44) {\rm GeV}$, and $m_{\Delta} = 0.658(49) {\rm GeV}$, respectively under the $\vec{A} = 0$ projection.
The overall decrease of hadron masses suggests a constituent quark with smaller mass of $m_{\mathrm{Q}} \simeq 0.2 {\rm GeV}$.

Interestingly, we find that nucleon and delta baryon masses are almost degenerate under the $\vec{A} = 0$ projection.
Here, we consider the physical meaning of the $\vec{A} = 0$ projection.
Under the $\vec{A} = 0$ projection, 
while color-electric fields can remain, 
color-magnetic fields 
$H_i^a=\epsilon_{ijk}(\partial_j A_k^a
-\partial_k A_j^a
-g f^{abc}A_j^b A_k^c)$
are inevitably zero, and therefore 
color-magnetic interactions vanish.
Since only the color-magnetic interactions are spin-dependent for static systems in QCD, 
mass splitting between different spin states are universally expected to vanish or to decrease under the $\vec{A} = 0$ projection.
This correspondence between 
color-magnetic interactions and 
spin-dependent hadron mass splitting 
seems consistent with the standard quark-model picture, 
since the $\mathrm{N}-\Delta$ mass splitting is generated 
by the color-magnetic interactions in the quark model. 

\section{A generalization of the $\vec{A} = 0$ 
projection in the Coulomb gauge} \label{sec:eigenmodeprojection}

To investigate the role of spatial gluons in the Coulomb gauge quantitatively, we generalize the $\vec{A} = 0$ projection so as to eliminate the spatial gluons  contribution continuously.
To this end, we utilize the Faddeev-Popov (FP) 
operator $M$ and consider expansion of the spatial gluon fields in terms of its eigenmodes. 
%
Roughly, the FP operator defined by Eq.~(\ref{eq:continuumFP}) 
resembles the Laplacian in the Coulomb gauge, 
where spatial gluon amplitude is fairly suppressed, 
and low-lying FP modes correspond to low-energy components.
In lattice QCD, the FP operator $M$ 
is expressed as a $8L_s^3 \times 8L_s^3$ matrix $M_{IJ}=M_{x,y}^{a,b}$ in Eq.~(\ref{eq:FPlattice}), 
with the index $I=(x,a)$ and $J=(y,b)$.

Since the FP operator is a real symmetric matrix, its eigenmodes $\psi_n$ satisfying 
\begin{equation}
M \psi_n = \lambda_n \psi_n \quad (n = 1, 2, \dots, 8L_s^3)
\end{equation}
can be taken real and normalized eigenvectors $\psi_n(t)$ at each time $t$ form a complete set 
\begin{equation}
\sum_{n}\psi_{n,s}^a(t)~ \psi_{n,s^{\prime}}^{a^{\prime}}(t) = \delta_{s,s^\prime} ~\delta^{a, a^{\prime}}.
\end{equation} 
With these eigenvectors, spatial gluon fields at each time $t$ can be uniquely expanded as
\begin{align}
A_{s,i}^a(t) = \sum_{n = 1}^{8 L_s^3} c_{n,i}(t) ~\psi_{n,s}^a(t), \qquad
c_{n,i}(t) \coloneqq \sum_{s,a} A_{s,i}^a(t) ~\psi_{n,s}^a(t).
\end{align}
Among the $8L_s^3$ eigenmodes, we consider a restriction of these modes, and define FP eigenmode projection 
for spatial gluon fields $A_{s,i}(t)$ 
as a replacement by 
\begin{align}
A_{s,i}^{a,\mathrm{projected}}(t)
=\sum_{n \in S} c_{n,i}(t)~\psi_{n,s}^a(t)
\end{align}
in the Coulomb gauge.
Here, $S$ denotes a subset of the whole set $\{1,\dots, 8L_s^3\}$. 
In this projection, the temporal gluon $A_4$ 
is unchanged. 
The eigenmode projection with whole set $S = \{1, \dots, 8L_s^3\}$ gives the original gauge configuration, whereas 
the projection with no eigenmode ($S = \phi$) becomes the  $\vec{A} = 0$ projection.
Then, we can connect the $\vec{A} = 0$ projected configuration with the original one by smoothly changing the subset $S$. 

In this study, we only consider eigenmode projections with $N_{\mathrm{low}}$ low-lying or $N_{\mathrm{high}}$ high-lying eigenmodes.
Specifically, we set
\begin{align}
S_{\mathrm{low}} = \{1, \dots, N_{\mathrm{low}}\}, \qquad
S_{\mathrm{high}} = \{8L_s^3 - N_{\mathrm{high}} + 1, \dots, 8L_s^3\}
\end{align}
for eigenmodes which are ordered as
$\lambda_1 \le \lambda_2 \le \dots \le \lambda_{8L_s^3}$.
For each gauge configuration and time $t$, we calculate 1638 low-lying and high-lying eigenmodes for the FP operator by using the Krylov-Schur solver implemented in SLEPc~\cite{Hernandez:2005:SSF}.
It is found that the 1638-th smallest eigenvalue corresponds to the low-lying eigenmode of about $(2 {\rm GeV})^2$.

Note that, unlike the Fourier expansion~\cite{Yamamoto:2008am,Yamamoto:2008ze}, 
the current expansion has no practical eigenvalue degeneracy except for the trivial eight zero-modes, 
and one can smoothly connect the $\vec{A} = 0$ projected gauge configuration with the original one.

\section{hadron and glueball masses under the eigenmode projection} \label{sec:hadron_mass}
\begin{figure*}[htb]
\includegraphics[width=\linewidth]{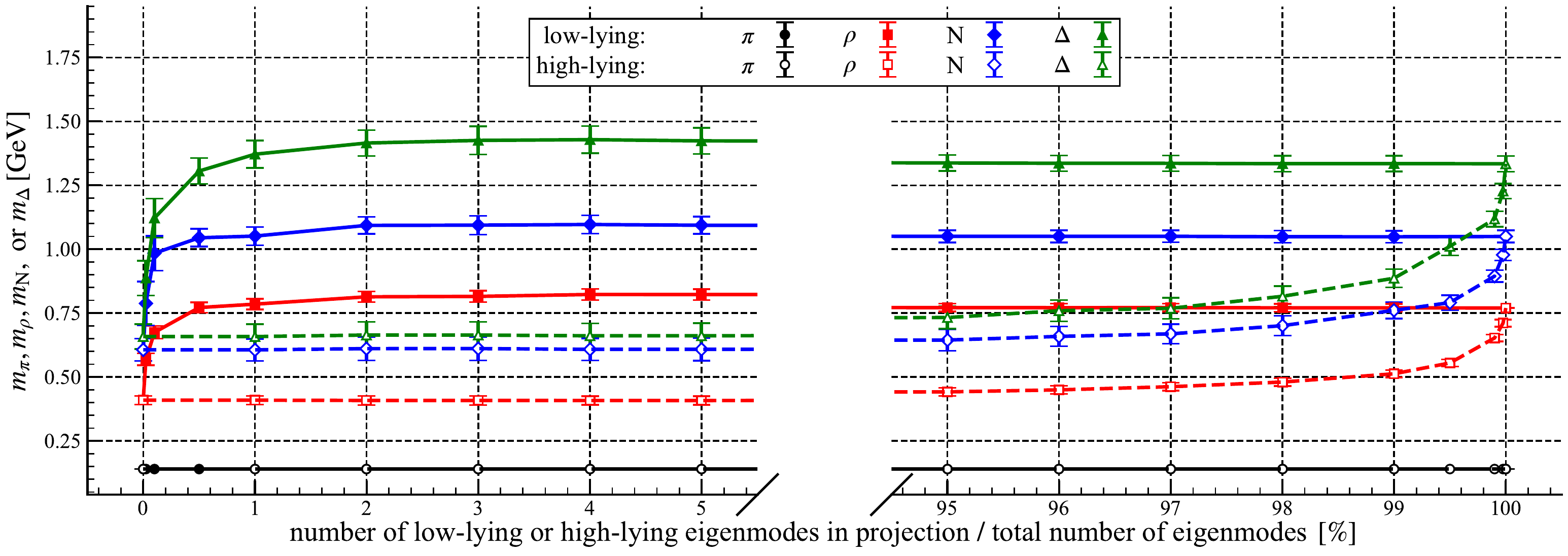}
\caption{\label{fig:hadron_mass}
Hadron masses under the eigenmode projection in GeV unit plotted against the number of low-lying or high-lying FP eigenmodes in the projection. 
Pion mass $m_\pi = 0.14 {\rm GeV}$ is used to set the physical hopping parameter for each measurement. 
The lattice spacing $a = 0.1051(15) {\rm fm}$ is used to set the scale.
The solid and dashed lines are connecting adjacent data points for visibility. 
}
\vspace{-0.5cm}
\end{figure*}

We calculate light hadron masses under the eigenmode projection.
Figure~\ref{fig:hadron_masses}~(c) and (d) show light hadron masses under the eigenmodes projection with 5\% low-lying and with 95\% high-lying eigenmodes, respectively.
Figure~\ref{fig:hadron_mass} shows 
hadron masses under the eigenmode projection plotted against the number of low-lying or high-lying FP eigenmodes. 
Some of these results are listed in Table~\ref{table:hadron_mass}.
\begin{table}[htb]
\caption{
Hadron masses under the eigenmode projection with low-lying FP eigenmodes.
}
\begin{center}
\begin{tabular}{cccc}
\hline 
num. of low-lying FP modes & $m_{\rho}$ [GeV] & $m_{\mathrm{N}}$ [GeV] & $m_{\Delta}$ [GeV] \\
\hline
0 (0 \%)          & 0.409(17)    & 0.607(44) & 0.658(49) \\ 
8 (0.02 \%)       & 0.569(23)    & 0.788(86) & 0.887(68) \\
33 (0.10 \%)      & 0.675(25)    & 0.984(68) & 1.122(77) \\
164 (0.50 \%)     & 0.772(20)    & 1.045(35) & 1.306(50) \\
328 (1.00 \%)     & 0.785(21)    & 1.051(36) & 1.372(54) \\
$8L_s^3$ (100 \%) & 0.77         & 1.050(24) & 1.334(31) \\
\hline
\end{tabular}
\end{center}
\label{table:hadron_mass}
\end{table}

We find monotonically increasing behaviors in Fig.~\ref{fig:hadron_mass}. 
The original light hadron masses are approximately reproduced with only one percent of low-lying eigenmodes.
Interestingly, Figs.~\ref{fig:hadron_masses} (c) and (d) are respectively very similar 
with Figs.~\ref{fig:hadron_masses} (a) and (b) in Sec.~\ref{sec:A0projection}, 
except for the values of hopping parameter $\kappa$.
These results indicate an important role of low-lying eigenmodes on light hadron masses.
From Fig.~\ref{fig:hadron_mass} and Table~\ref{table:hadron_mass}, 
we also find that the $\mathrm{N}-\Delta$ mass splitting becomes evident when projected with more than 0.1 \% low-lying eigenmodes.

Finally, to go beyond the quark model, 
we investigate $0^{++}$ and $2^{++}$ glueball masses 
in the FP eigenmode projection in the Coulomb gauge. 
The mass splitting of $0^{++}$ and $2^{++}$ glueballs 
are also considered to be due to the color-magnetic interaction. 

\begin{figure*}[htb]
\begin{center}
\includegraphics[width=1.0\linewidth]{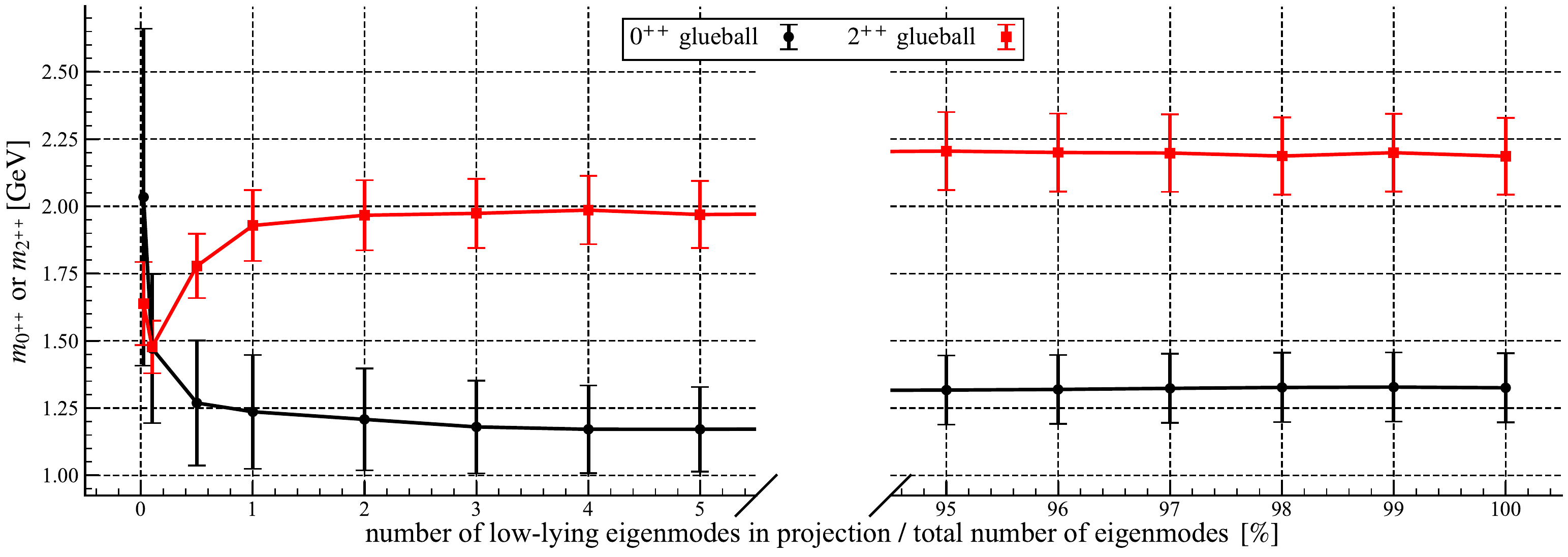}
\end{center}
\vspace{-0.25cm}
\caption{\label{fig:glueball_mass}
Glueball masses under the eigenmode projection in GeV unit plotted against the number of low-lying FP eigenmodes in the projection.
}
\end{figure*}

These glueball masses can be extracted from correlators
\begin{align}
C(t) = \phi(t) \phi(0), \qquad 
\phi(t) = 
\begin{cases}
\sum_{\vec{x}} {\rm Re}~{\rm tr}(P_{12} + P_{23} + P_{31})(\vec{x}, t) & \text{for $0^{++}$}, \\
\sum_{\vec{x}} {\rm Re}~{\rm tr}(P_{12} - P_{13})(\vec{x}, t) & \text{for $2^{++}$},
\end{cases}
\label{eq:glueball_corr}
\end{align}
where $P_{ij}$ is the plaquette in $i-j$ plane and the vacuum contribution must be subtracted from the correlator for $0^{++}$ glueball mass~\cite{Ishikawa:1982tb}.
For the glueball calculation, although 500 gauge configurations might not be enough statistically  \cite{Teper:1987wt}, 
we calculate approximate $0^{++}$ and $2^{++}$ glueball masses under the eigenmode projection with low-lying eigenmodes,
from single exponential fits to the correlators~(\ref{eq:glueball_corr}).
Those results are summarized in Fig.~\ref{fig:glueball_mass}.

\begin{figure*}[htb]
\begin{center}
\includegraphics[width=0.8\linewidth]{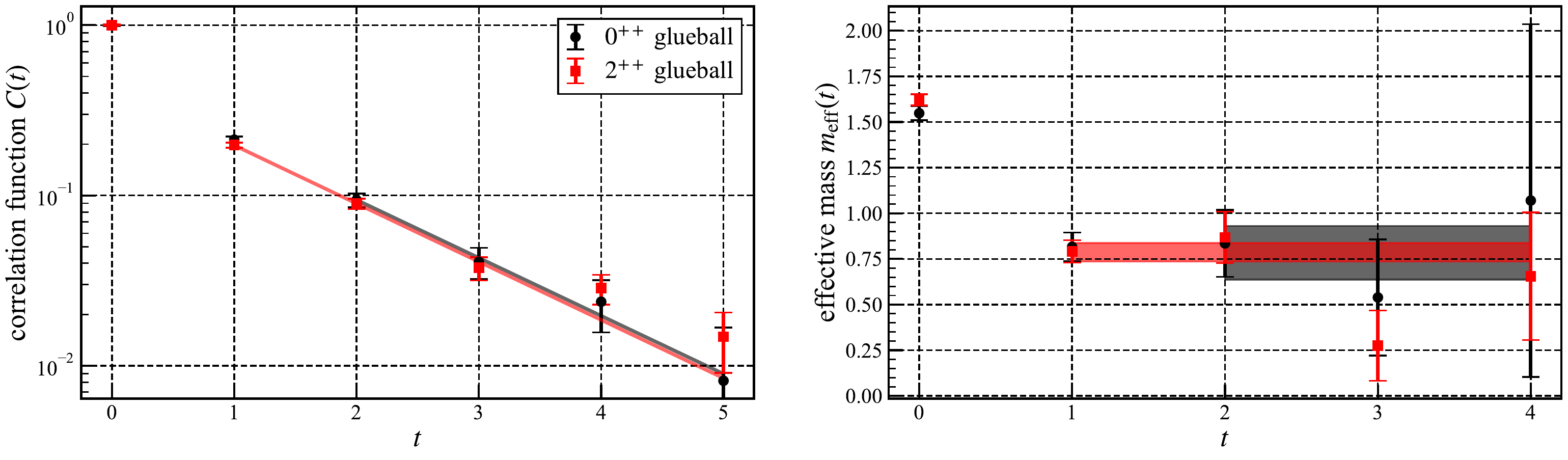}
\end{center}
\vspace{-0.5cm}
\caption{\label{fig:glueball_mass_only_small_33}
Glueball mass measurement under the eigenmode projection with 0.1 \% low-lying FP eigenmodes.
Correlation functions are normalized such that $C(0) = 1$.
}
\end{figure*}

We also show glueball mass mesurement under the eigenmode projection with 0.1 \% low-lying eigenmodes in Fig.~\ref{fig:glueball_mass_only_small_33}.
In the case of projection with high-lying eigenmodes, 
the correlation functions are too noisy that we can not extract the masses at all.

From Fig.~\ref{fig:glueball_mass}, it seems that $0^{++}$ and $2^{++}$ glueball masses are approximately degenerate near the $\vec{A} = 0$ projection.
This can be also seen from Fig.~\ref{fig:glueball_mass_only_small_33}, where correlation functions are very alike.
They approximately converge to the original values with just one percent of low-lying eigenmodes.
We also find that the $0^{++}-2^{++}$ glueball mass splitting becomes evident when projected with more than 0.1 \% low-lying eigenmodes.
Thus, we obtain similar results also in the case of glueball masses.

\section{Summary and Conclusion} \label{sec:summary}

In this paper, aiming at the relation between QCD and the quark model, we have investigated the role of spatial gluons for hadron masses in the Coulomb gauge. 
In the $\vec{A} = 0$ projection, 
where all the spatial gluon fields are set to zero, 
we have found that nucleon and delta baryon masses are almost degenerate. This suggests that 
the N-$\Delta$ mass difference arises from the color-magnetic interactions, which is consistent with the quark model picture.
Next, as a generalization of this projection, 
we expanded spatial gluon fields in terms of Faddeev-Popov eigenmodes 
and left only some partial components.
We have found that the original ${\rm N}-\Delta$ and $0^{++}-2^{++}$ glueball mass splittings are almost reproduced only with 1 \% low-lying components. 
This suggests that low-lying color-magnetic interaction leads to the hadron mass splitting. 

Our study suggests a possibility that the quark model can be derived from Coulomb gauge QCD by a reduction of high-lying spatial gluons above $1.3 {\rm GeV}$.
Also, a QCD-based constituent gluon model 
might be formulated from Coulomb gauge QCD with the reduction of high-lying spatial gluons, and its combination with the quark model might describe wide category of hadrons including glueballs and hybrids.

H.O. is supported by a Grant-in-Aid 
for JSPS Fellows (Grant No.21J20089).
H.S. is supported by a Grants-in-Aid for
Scientific Research [19K03869] from JSPS. 
This work was in part based on Bridge++ code~\cite{Ueda:2014rya}.
We used SLEPc~\cite{Hernandez:2005:SSF} to solve eigenvalue problems.
The numerical simulations was carried out on Yukawa-21 at YITP, Kyoto University.

\bibliography{conf}
\end{document}